\documentclass[amsfonts, amssymb, amsmath, reprint, showkeys, nofootinbib, prd, floatfix, superscriptaddress, natbib]{revtex4-1}
\usepackage[english]{babel}
\usepackage[utf8]{inputenc}
\usepackage[colorinlistoftodos, color=green!40, prependcaption]{todonotes}
\usepackage{multirow}
\usepackage{float}
\usepackage[pdftex, pdftitle={Article}, pdfauthor={Author}]{hyperref} 
\usepackage{amsthm}
\usepackage{mathtools}
\usepackage{physics}
\usepackage{xcolor}
\usepackage{graphicx}
\usepackage[left=23mm,right=13mm,top=35mm,columnsep=15pt]{geometry} 
\usepackage{adjustbox}
\usepackage{placeins}
\usepackage[T1]{fontenc}
\usepackage{lipsum}
\usepackage{csquotes}
\usepackage{sidecap}
\usepackage[normalem]{ulem}

\newcommand{\aap}{Astron. Astrophys.}

\newcommand{\apjs}{Astrophys. J. Suppl.}
\newcommand{\jcap}{J. Cosmol. Astropart. Phys.}
\newcommand{\mnras}{Mon. Not. Roy. Astron. Soc.}
\newcommand{\apss}{Astrophys. Space Sci.}
\newcommand{\araa}{Annu. Rev. Astron. Astrophys.}
\newcommand{\physrep}{Phys. Rep.}
\newcommand{\plb}{Phys. Lett. B}

\begin{document}

\title{Percent-level constraints on baryonic feedback with spectral distortion measurements}

\author{Leander Thiele}
\email{lthiele@princeton.edu}
\affiliation{Department of Physics, Princeton University,
             Princeton, NJ 08544, USA}

\author{Digvijay Wadekar}
\affiliation{School of Natural Sciences, Institute for Advanced Study,
             Princeton, NJ 08540, USA}

\author{J. Colin Hill}
\affiliation{Department of Physics, Columbia University,
             New York, NY 10027, USA}
\affiliation{Center for Computational Astrophysics, Flatiron Institute,
             New York, NY 10010, USA}

\author{Nicholas Battaglia}
\affiliation{Department of Astronomy, Cornell University,
             Ithaca, NY 14853, USA}

\author{Jens Chluba}
\affiliation{Jodrell Bank Centre for Astrophysics, Department of Physics and Astronomy,
             The University of Manchester, Manchester M13 9PL, UK}

\author{Francisco Villaescusa-Navarro}
\affiliation{Center for Computational Astrophysics, Flatiron Institute,
             New York, NY 10010, USA}
\affiliation{Department of Astrophysical Sciences, Princeton University,
             Princeton, NJ 10027, USA}

\author{Lars Hernquist}
\affiliation{Center for Astrophysics, Harvard \& Smithsonian,
             Cambridge, MA 02138, USA}

\author{Mark Vogelsberger}
\affiliation{Kavli Institute for Astrophysics and Space Research,
             Massachusetts Institute of Technology,
	     Cambridge, MA 02139, USA}

\author{Daniel Angl\'es-Alc\'azar}
\affiliation{Department of Physics, University of Connecticut,
             Storrs, CT 06269, USA}
\affiliation{Center for Computational Astrophysics, Flatiron Institute,
             New York, NY 10010, USA}

\author{Federico Marinacci}
\affiliation{Department of Physics and Astronomy ``Augusto Righi'',
             University of Bologna, Bologna, I-40129, Italy}

\begin{abstract}
High-significance measurements of the monopole thermal Sunyaev-Zel'dovich CMB spectral distortions
have the potential to tightly constrain poorly understood baryonic feedback processes.
The sky-averaged Compton-$y$ distortion and its relativistic correction are measures
of the total thermal energy in electrons in the observable universe
and their mean temperature.

We use the CAMELS suite of hydrodynamic simulations to explore possible constraints
on parameters describing the subgrid implementation of feedback from active galactic nuclei
and supernovae, assuming a PIXIE-like measurement.

The small $25\,h^{-1}\text{Mpc}$ CAMELS boxes present challenges due to significant sample variance.
We utilize machine learning to construct interpolators through the noisy simulation data.
Using the halo model, we translate the simulation halo mass functions into correction factors
to reduce sample variance where required.

Our results depend on the subgrid model.
In the case of IllustrisTNG, we find that the best-determined parameter combination
can be measured to $\simeq 2\,\%$ and corresponds to a product of AGN and SN feedback.
In the case of SIMBA, the tightest constraint is $\simeq 0.2\,\%$ on
a ratio between AGN and SN feedback.
A second orthogonal parameter combination can be measured to $\simeq 8\,\%$.
Our results demonstrate the significant constraining power a measurement of the late-time
spectral distortion monopoles would have for baryonic feedback models.
\end{abstract}

\maketitle

\section{Introduction}

A robust prediction of the standard model of cosmology is the presence of subtle deviations
from the blackbody spectrum in the cosmic microwave background (CMB) \citep[e.g.,][]{Sunyaev2013, Chluba2016}.
Such monopole spectral distortions can be caused by a variety of physical processes in both the early-
and late-time Universe.
Among these various distortion signals,
most easily accessible to near-future experiments will be the $y$-type distortion.\footnote{For conciseness,
we will use the term ``$y$ distortion'' or similar to mean both the non-relativistic and the relativistic effects.}
The $y$ distortion arises from inverse Compton scattering of CMB photons with hot electrons,
the thermal Sunyaev-Zel'dovich (tSZ) effect \cite{Zeldovich1969,Sunyaev1970}.
The $y$ signal is mostly sourced by massive, collapsed structures at $z \lesssim 2$.
Thus, the tSZ effect informs us about two aspects:
First, the thermodynamic properties of the hot electron gas in massive halos;
these are very sensitive to astrophysical small-scale processes.
Second, the abundance of such halos, which translates into a constraint on
the amount and clustering of matter in the Universe.
This makes SZ cluster measurements a unique tool for cosmology and astrophysics \citep{Carlstrom2002, SZreview2019}.

\begin{figure}
\includegraphics[width=0.5\textwidth]{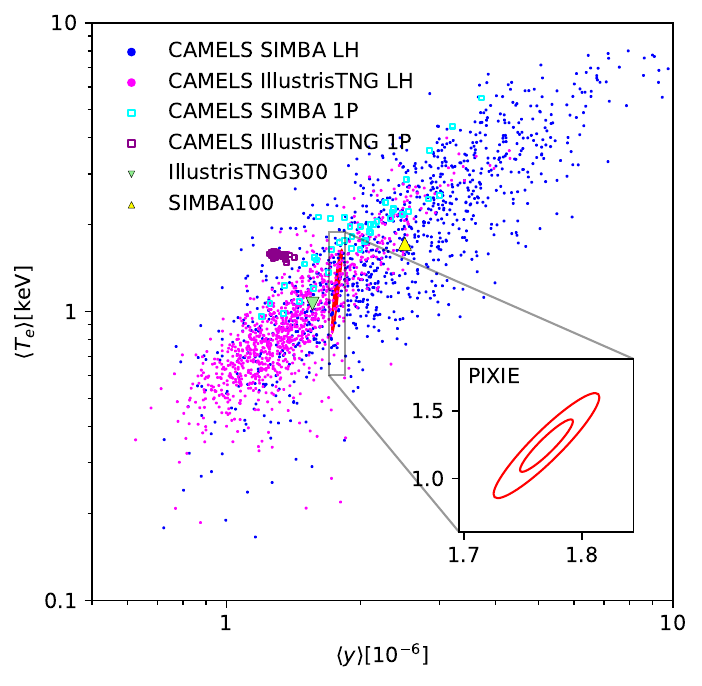}
\caption{Comparison between the range of CAMELS simulations with the forecast covariance
         matrix for the PIXIE experiment
	 (the PIXIE ellipses are centered at the fiducial model assumed in Ref.~\cite{Abitbol2017}:
	 $\langle y \rangle_\text{fid}=1.77 \times 10^{-6}$,
	 $\langle T_e \rangle_\text{fid}=1.24\,\text{keV}$;
	 they correspond to $68$ and $95\,\%$ CL).
	 From CAMELS, we plot both the LH and the 1P set. The LH data points are contaminated
	 by sample variance in addition to varying subgrid parameters,
	 while the 1P points have the same initial conditions and only differ by their subgrid parameters.
	 We have also indicated the values measured from the large boxes (triangles),
	 as discussed in Sec.~\ref{sec:sims:lb}.
	 All simulation data have been rescaled to the fiducial CAMELS cosmology but
	 each data point corresponds to a different subgrid model.
	 It is worth noting that the 1P initial conditions appear to be slightly atypical,
	 with generally larger $\langle T_e \rangle$ than expected from the LH set.}
\label{fig:PIXIEcov}
\end{figure}

To illustrate this point, consider Fig.~\ref{fig:PIXIEcov}.
There, we plot a range of theory predictions (blue and magenta points)
from the CAMELS suite of hydrodynamic simulations \cite{FVN2021}.
The axes are the two distortion monopole observables considered in this work, further discussed below.
Between the plotted simulations the astrophysical sub-grid model differs,
translating into a large spread of theory predictions.\footnote{For IllustrisTNG,
the spread is dominated by sample variance (as can be seen by comparing the LH and 1P points),
a point that we will return to later.}
For comparison, we also show a covariance matrix for a near-future monopole distortion measurement
(c.f.\ Sec.~\ref{sec:PIXIE}).
The forecast measurement errors are tiny compared to the current theoretical uncertainty,
which means that a near-future measurement would provide a substantial gain in information on astrophysics.
This basic observation forms the underpinning of the calculations performed in the following:
forecast constraints on simulation subgrid models from a measurement of the $y$ distortion monopoles,
using simulations from the CAMELS suite.

Conventionally, the $y$-distortion is separated into a non-relativistic
and a relativistic \citep{Sazonov1998,Challinor1998,Itoh1998,Chluba2005,Nozawa2006} component,
each having a distinct spectral signature that makes it possible to disentangle
them observationally.\footnote{All these signals can be accurately modeled using {\tt SZpack} \citep{Chluba2012SZpack}.}
The non-relativistic contribution is determined by a line-of-sight integral over
electron pressure,\footnote{We set the speed of light and Boltzmann's constant to one.}
\begin{equation}
\langle y \rangle \equiv \langle y(\hat n) \rangle_{\hat n}
  = \int \frac{d\hat n}{4\pi} \frac{\sigma_\text{T}}{m_e} \int P_e(\hat n, l)\,dl\,,
\label{eq:ydef}
\end{equation}
where $\sigma_\text{T}$ denotes the Thomson cross section, $m_e$ is electron mass,
$\hat n$ is the line of sight, and $l$ is physical length along the line of sight.
To leading order, the relativistic component is proportional to the $y$-weighted mean electron temperature~\cite{Hill2015}
\begin{equation}
\langle T_e \rangle \equiv \langle T_e(\hat n) \rangle_{\hat n}
  = \langle y \rangle^{-1} \int \frac{d\hat n}{4\pi} \frac{\sigma_\text{T}}{m_e} \int [T_e P_e](\hat n, l)\,dl\,.
\label{eq:Tdef}
\end{equation}
This effective temperature is typically higher than the mass-weighted (or $\tau$-weighted) temperature \citep{Kay2008, Lee2020},
and can be directly obtained from a moment expansion of the SZ signal
(see Appendix~\ref{app:rSZ-weighting} for additional discussion).

In this work we focus on the dominant and theoretically well-established
contributions to the distortion signals.
In particular, we will neglect the $\simeq 10\,\%$ contribution of reionization to $\langle y \rangle$,
which we discuss in Appendix~\ref{app:reio},
as well as signals due to the Milky Way and Local Group (which are estimated to be another one and two
orders of magnitude below the reionization signal, respectively).
We also neglect other, more exotic sources of $y$-distortions,
for example from primordial magnetic field heating \citep[e.g.,][]{Jedamzik2000, Kunze2014,Chluba2015}
or decaying particles \citep[e.g.,][]{Sarkar1984,Hu1993b, Chluba2013fore,HaiChlKam15,BolChl20,Ali21}.
Conversely, if one is interested in using the $y$-distortions to constrain or detect
such processes beyond standard $\Lambda$CDM,
astrophysical feedback must be very well understood.

There are further relativistic corrections involving higher moments of the electron temperature,
as well as the kinetic Sunyaev-Zel'dovich effect sourced by coherent motion;
we will neglect these complications and simply treat $\langle y \rangle$ and $\langle T_e \rangle$ as
observables, as in Ref.~\cite{Hill2015}.

\begin{figure}
\includegraphics[width=0.48\textwidth]{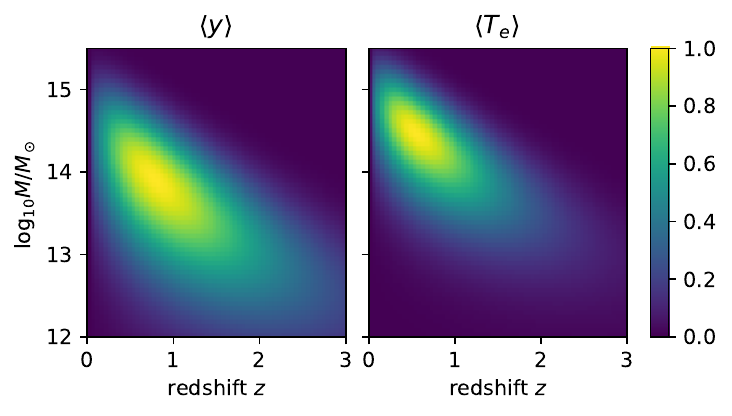}
\caption{Illustration of the halo mass and redshift contributions to the $y$ distortion
         observables.
	 Plotted is the normalized integrand in Eq.~\eqref{eq:hmintegral}, using the fitting formulae
	 described in Sec.~\ref{sec:nn:hmf}.}
\label{fig:Mz}
\end{figure}

Locally, the tSZ effect is well-established observationally as a CMB temperature change correlating
with the locations of clusters.
However, the global distortion to the CMB spectrum has not yet been detected.
Only an upper limit on the non-relativistic $y$ distortion exists from the COBE FIRAS experiment,
which yielded $|\langle y \rangle| < 15 \times 10^{-6} \, (95\,\%\,\text{cl})$ \cite{Fixsen1996},
which is about one order of magnitude above the expected $\Lambda$CDM signal~\cite{Hill2015}.
We thus anticipate significant detections with future CMB spectroscopy \citep{Kogut2011,Kogut2016SPIE, BISOU2021, Chluba2021}.

A monopole measurement would be complementary to the existing higher-moment 
tSZ analyses~\citep[e.g.,][]{Plagge2010,Hand2011,PlanckStkLBG,Planck2013pressureprofiles,
Hill2014,Greco2015,Planck2016tSZmap,Vikram2017,
Dietrich2019,Madhavacheril2020,Bleem2021,Schaan2021,Amodeo2021,Vavagiakis2021,
Pratt2021},
since it features very different systematics and would also yield the relativistic component
at high significance \cite{Abitbol2017}.
Furthermore, in contrast to cluster-stacking and power spectrum \citep{Hurier2017rSZ,Erler2017,Remazeilles2020} approaches,
the $\langle y \rangle$ measurement is more sensitive to lower-mass objects, as illustrated in Fig.~\ref{fig:Mz}.

The $\langle y \rangle$ and $\langle T_e \rangle$ signals constitute unique probes of baryonic
physics in galaxy clusters and groups.
Since $\langle y \rangle$ probes thermal energy, it is subject to the energy conservation equation
\begin{equation}
\underbrace{ E^\text{th,tot}_e}_\text{Total} = \underbrace{ E^\text{coll}_e}_\text{Collapse}
                                               + \underbrace{ E^\text{inj}_e}_\text{Injected}
                       		               - \underbrace{ E^\text{cool}_e}_\text{Cooling}\,.
\end{equation}
The most uncertain term in the above equation is $E^\text{inj}_e$ which can largely be attributed
to feedback processes from massive stars, supernovae (SNe), and active galactic nuclei (AGN).
These processes inject additional energy into the interstellar, intergalactic, and intracluster
media (ISM, IGM, and ICM, respectively).
Such feedback processes are standard ingredients in any theoretical models of galaxy formation,
both semi-analytic \citep[e.g.,][]{Croton2006,Somerville2015}
and simulation-based models
\citep[e.g.,][]{Springel2005,Schaye2015,DAA2017,Kaviraj2017,Hopkins2018,Vogelsberger2020}.

The most reliable way to explore how feedback models influence the $y$-distortions
is by analyzing hydrodynamical simulations with qualitatively and quantitatively different subgrid prescriptions.
Such an approach is complicated by the fact that, owing to their bias towards rare high-density
peaks, the distortion signals are heavily influenced by sample variance.
Furthermore, the parameter space of subgrid models is vast and poorly explored,
meaning that ideally we would need many large-volume hydrodynamical simulations,
which is currently not feasible.
We will demonstrate later that these problems can be overcome by utilizing machine learning methods
as well as analytical corrections using the halo model.

Since the $y$-distortions are predominantly sourced by galaxy groups and clusters \cite{Hill2015},
an analytical description based on the halo model \citep{Peebles1965,PressSchechter1974,Cooray2002}
is a natural first approximation.
In the halo model formalism, we assume spherically symmetric halos described only by mass $M$
and redshift $z$, yielding
\begin{equation}
\langle y \rangle_\text{hm} = \frac{\sigma_\text{T}}{m_e} \int dz dM
    \frac{(1+z)^2}{4\pi H(z)} \frac{dn}{dM} \int d\vec r\,P_e(|\vec r|; M,z)\,,
\label{eq:hmintegral}
\end{equation}
where $dn/dM$ is the halo mass function and $\vec r$ denotes position within a given halo,
and the expression assumes a flat universe.
The expression for $\langle T_e \rangle_\text{hm}$ is analogous. 
Note that the halo model neglects the IGM contribution discussed in Appendix~\ref{app:IGM}.

In the following, it will be useful to think of the observables
$x_i \equiv \{\langle y \rangle, \langle T_e \rangle\}$
in terms of the approximate factorization
\begin{equation}
x_i \sim f_i^c(\sigma_8, \Omega_m, ...) f_i^b(\{A_j\}) f_i^\text{CV}(\delta)\,,
\label{eq:factorization}
\end{equation}
where $f_i^c$ describe the dependence on cosmological parameters,
$f_i^b$ are functions of a set of feedback parameters $A_j$,
and $f_i^\text{CV}$ depends on the initial conditions and thus encapsulates sample variance.
Such factorizations are frequently-used and good approximations to observables that are well-described
by the halo model.
Nonetheless, our results typically only weakly depend on the validity of this approximation.

It should be noted that we group all our uncertainty on the simulation sub-grid model in
the feedback parameters $A_j$.
These supernova and AGN feedback parameters, further elaborated on in Sec.~\ref{sec:sims:CAMELS},
predominantly affect the ICM contribution to the distortion signals.
There is a non-negligible IGM contribution, however, which in Appendix~\ref{app:IGM}
we show is a $\sim 10\,\%$ effect with $\sim 40\,\%$ theoretical uncertainty.

The rest of this paper is structured as follows.
In Sec.~\ref{sec:sims} we describe the CAMELS simulations as well as the larger reference
boxes.
Sec.~\ref{sec:PIXIE} provides a short summary of the assumed experimental setup used for forecasting.
In Sec.~\ref{sec:nn} we describe how we interpolate through the CAMELS data.
Sec.~\ref{sec:results} contains the main results of this work, namely, dependence
of the distortion signals on feedback parameters and a Fisher forecast.
We conclude in Sec.~\ref{sec:concl}.
The appendices contain several technical details and some new computations that did
not fit in the main discussion.

\section{Simulations}
\label{sec:sims}

The $y$-distortion components described above can easily be measured from a hydrodynamical
simulation. In fact, Eq.~\eqref{eq:ydef} can be rewritten as
\begin{equation}
\langle y \rangle = \frac{\sigma_\text{T}}{m_e} \int dz \frac{(1+z)^2}{H(z)} \langle P_e^c(z) \rangle\,,
\label{eq:simintegral}
\end{equation}
where $P_e^c$ is now in comoving units and the average is over the volume of a given simulation
snapshot.
An analogous expression holds for $\langle T_e \rangle$.

\subsection{CAMELS}
\label{sec:sims:CAMELS}

We primarily use the CAMELS suite of hydrodynamical simulations \cite{FVN2021} that consists of several thousand $25\,h^{-1}\text{Mpc}$ boxes,
each run with $256^3$ dark matter particles and $256^3$ initial fluid elements.
Each simulation is described by the following parameters:
\textsl{i}) the simulation code/subgrid model,
\textsl{ii}) two cosmological parameters ($\sigma_8$, $\Omega_m$),
\textsl{iii}) four feedback parameters\footnote{Note that other works, e.g.\ Ref.~\cite{Wadekar2022}, explicitly differentiate
in their notation between the IllustrisTNG and SIMBA feedback parameters. We choose not to do so since it simplifies the notation
in various places.} ($A_\text{SN1}$, $A_\text{SN2}$, $A_\text{AGN1}$, $A_\text{AGN2}$),
and \textsl{iv}) the random seed for the initial conditions.
The remaining cosmological parameters are fixed at \textsl{Planck}-compatible flat $\Lambda$CDM values,
$\Omega_b=0.049$, $h=0.6711$, $n_s=0.9624$ \cite{Planck2018params}.

Two different simulation codes are used.
The simulations labeled `IllustrisTNG' were run with the Arepo code \cite{Springel2005,Weinberger2020} and the same
subgrid model as the flagship IllustrisTNG simulations \cite{IllustrisTNG1,IllustrisTNG2,IllustrisTNG3,IllustrisTNG4,IllustrisTNG5,Nelson2019}.
The simulations labeled `SIMBA' were run with the GIZMO code \cite{Hopkins2015} and the same
subgrid model as the flagship SIMBA simulations \cite{Dave2019}.
These codes differ substantially in their subgrid implementations, so having comparable
simulations with both gives a good indication of the theoretical uncertainty.

The cosmological parameters $\sigma_8$, $\Omega_m$ are varied in the intervals $[0.6,1.0]$ and $[0.1, 0.5]$,
respectively, the fiducial model being $(0.8, 0.3)$.

The precise definition of the four feedback parameters is given in Ref.~\cite{FVN2021}.
Broadly, $A_{\text{SN}i}$ parameterize the subgrid prescription for galactic winds,
while $A_{\text{AGN}i}$ describe the efficiency of black hole feedback.
The $i=1$ components can be thought of `energy' normalizations,
while the $i=2$ components scale the speed of outflows.
In detail, however, the meaning of the feedback parameters differs substantially
between IllustrisTNG and SIMBA.
We thus caution against any direct comparison in terms of the feedback parameters
between the two subgrid models.
All the $A_j$ are multiplicative factors relative to the fiducial efficiencies
in the original IllustrisTNG and SIMBA simulations;
the fiducial model is therefore unity for all feedback parameters.
$A_\text{SN1}$ and $A_\text{AGN1}$ are varied in $[0.25, 4.0]$,
while $A_\text{SN2}$ and $A_\text{AGN2}$ are varied in $[0.5, 2.0]$.
These intervals were chosen heuristically by the CAMELS team, and as we will see
the corresponding variations in the $y$ observables differ drastically between IllustrisTNG and SIMBA.
Some of the simulations at the corners of parameter space are so extreme that they are certainly not
realistic, for example with regard to galaxy properties.

For each of the two simulation codes, the CAMELS suite comprises the following sets of simulations,
all of which will be used in this work.
\begin{itemize}
\item LH: (latin hypercube) 1000 simulations in which cosmology and feedback parameters
          are varied on a latin hypercube, each run having a different random seed,
\item 1P: (one parameter at a time) 10 variations for each cosmological and feedback parameter
          individually at fixed random seed,
\item CV: (cosmic variance) 27 simulations at the fiducial model with differing random seeds.
\end{itemize}

\subsection{Larger boxes}
\label{sec:sims:lb}

In addition to the CAMELS suite, we also use larger boxes, namely
IllustrisTNG300-1 ($205\,h^{-1}\text{Mpc}$) and SIMBA100 ($100\,h^{-1}\text{Mpc}$).
These simulations are useful to calibrate against the fact that
$\langle f_i^\text{CV}(\delta) \rangle$ (c.f. Eq.~\ref{eq:factorization}) is relatively more biased for the small $25\,h^{-1}\text{Mpc}$ CAMELS boxes.
Denoting an observable measured in one of the large boxes as $x_i^\text{lb}$ (c.f. Eq.~\eqref{eq:factorization} for the notation),
we compute an estimator for this multiplicative bias as
\begin{equation}
b_i = x_i^\text{lb}/\langle x_i \rangle_\text{CV}\,,
\label{eq:bias}
\end{equation}
where the latter average is over the CAMELS CV set and we apply the correction formula
\begin{equation}
x_i \leftarrow x_i b_i
\end{equation}
to the $\langle y \rangle$, $\langle T_e \rangle$ measured in the CAMELS simulations.
Values for the $b_i$ are listed in Tab.~\ref{tab:compCVlb}.

The large boxes have the same subgrid prescription as the CAMELS fiducial model
with the corresponding label, but slightly different cosmologies.
We account for this fact by rescaling $x_i^\text{lb}$ to the CAMELS fiducial cosmology,
using power laws in $h$, $\Omega_m$, $\Omega_b$, $n_s$, $\sigma_8$.
These power laws were fitted using the halo model, since the scalings that could be derived
from the 1P set are unreliable and do not encompass all differences in cosmology.
We refer to Sec.~\ref{sec:nn:hmf} for a detailed description of the assumptions
made in the halo model calculation.
For reference, the power laws are listed in Appendix~\ref{app:pwrlaws}.

\begin{table*}
\centering
\begin{tabular}{ccccccccc}
           & \multicolumn{2}{c}{$\log_{10} 10^6\langle y \rangle$}
	   & \multicolumn{2}{c}{$\log_{10} \langle T_e \rangle [\text{keV}]$}
           & \multicolumn{2}{c}{$10^6\langle y \rangle$}
	   & \multicolumn{2}{c}{$\langle T_e \rangle [\text{keV}]$} \\
	   & IllustrisTNG       & SIMBA              & IllustrisTNG       & SIMBA
	   & IllustrisTNG       & SIMBA              & IllustrisTNG       & SIMBA                \\
\hline
CAMELS CV  & 0.046 $\pm$ 0.026  & 0.369 $\pm$ 0.014  & -0.386 $\pm$ 0.034 & 0.1407 $\pm$ 0.0054
           & 1.11               & 2.34               & 0.41               & 1.38                 \\
large box  & 0.1928             & 0.3995             & 0.02728            & 0.2349
           & 1.56               & 2.51               & 1.06               & 1.72                 \\
$\Delta [\sigma]$
           & -5.6               & -2.1               & -12                & 17
	   & & & & \\
$b_i$      & & & &
           & 1.41               & 1.07               & 2.59               & 1.25
\end{tabular}
\caption{Comparison between statistics over the CAMELS CV set and the larger boxes
         of sizes $L=205$ and $100\,h^{-1}\text{Mpc}$ for IllustrisTNG and SIMBA respectively.
	 The left half of the table is in $\log$ since log-normal is a decent approximation;
	 for clarity the right half gives the same information in more familiar units but without
	 error bars.
	 The error bars in the first line are on the mean of the CV set.
	 The error bars on the large box measurements are, in the naive Poissonian approximation
	 $\sqrt{(L/25\,h^{-1}\text{Mpc})^3/27}$, smaller by factors
	 4.5 and 1.5 for IllustrisTNG and SIMBA respectively, and omitted in the table.
	 All large-box measurements have been rescaled to the fiducial CAMELS cosmology,
	 as explained in Sec.~\ref{sec:sims:lb}.
	 The third line lists the differences between the large-box and CV measurements, in units
	 of the error bars on the CV means; these numbers highlight significant biases in the CV
	 set due to box-size effects.
	 The last line gives the corresponding bias factors introduced in Eq.~\eqref{eq:bias}.}
\label{tab:compCVlb}
\end{table*}

There is, of course, a remaining bias due to the finite volume of both the larger boxes
and the CV set.
However, as seen in Tab.~\ref{tab:compCVlb}, this is a relatively small source of error compared
to the overall bias, since Eq.~\eqref{eq:bias} is significantly different from unity for all cases
except perhaps the SIMBA $\langle y \rangle$.
We note that the finite error bars on the large box measurements do not affect our forecasts
in the later parts of this work.

\section{PIXIE experimental model}
\label{sec:PIXIE}

For deriving our forecasts later in section~\ref{sec:results},
we use parameters corresponding to the `extended' PIXIE experiment \citep{Kogut2011,Kogut2016SPIE,Kogut2020}
from Ref.~\cite{Abitbol2017}.
Their most complete model includes marginalization over a variety of foregrounds,
namely galactic dust thermal emission, the cosmic infrared background, 
synchrotron radiation, free-free emission, spinning dust (anomalous microwave emission),
and integrated CO.
For each foreground component, fitting formulas were assumed for the spectral energy
distribution (SED) according to \textit{Planck} measurements~\cite{Planck2016foregrounds}.
There are some caveats to this approach, related to the spatial variation of the foregrounds
and the relatively simplistic modeling of their SEDs \citep[see][for related discussion]{Rotti2021}.
However, for the purposes of this work, the derived forecasts should be accurate enough.
The forecast considers all non-negligible CMB spectral distortion signals (blackbody, $y$, relativistic $y$, $\mu$)
and marginalizes over them when computing the $\langle y \rangle$-$\langle T_e \rangle$ posterior.

While Ref.~\cite{Abitbol2017} used a full MCMC pipeline to arrive at their posteriors,
in the $\langle y \rangle$-$\langle T_e \rangle$ plane, these are well approximated as Gaussian.
We thus compress the marginalized posterior into a simple $2 \times 2$ covariance matrix,
illustrated in the inset in Fig.~\ref{fig:PIXIEcov}.
In general, the fiducial models used in this work differ somewhat from the fiducial model
of Ref.~\cite{Abitbol2017};
we assume that the covariance matrix is constant regardless of the mean.
This approximation is not a dominant source of systematic uncertainty.

\section{Interpolating neural networks}
\label{sec:nn}

Using Eq.~\eqref{eq:simintegral}, we compute $\langle y \rangle$ and $\langle T_e \rangle$ for
the CAMELS LH, CV, and 1P sets.
It is our primary goal to extract the functions $f_i^b(A_\text{SN1}, A_\text{AGN1}, A_\text{SN2}, A_\text{AGN2})$,
in the language of Eq.~\eqref{eq:factorization}.
It may be argued that in the limit in which these $f_i^b$ factorize even further into functions of
the individual feedback parameters (which, as we shall see, is quite a good approximation),
the 1P set should be all we need.
However, the substantial sample variance in the small CAMELS boxes makes this approach unreliable.
Therefore, we will instead utilize the LH set.
There, $\sigma_8$, $\Omega_m$, and the four feedback parameters are varied by sampling from a latin hypercube.
Furthermore, each data point is a noisy sample in a 6-dimensional space.
Thus, we will use neural networks as smoothing interpolators through the LH set.

\subsection{Training}
\label{sec:nn:training}

We aim to learn functions 
\begin{equation}
\langle y \rangle\ \mathrm{or}\ \langle T_e \rangle = F_\alpha(\sigma_8,\Omega_m,\{A_j\})\, ,
\end{equation}
where $\alpha \in \{\text{IllustrisTNG}, \text{SIMBA}\}$
and $F$ is parameterized as a multi-layer perceptron.
A multi-layer perceptron is a series of affine transformations $W\mathbf{x}+\mathbf{b}$, each
followed by a non-linear activation function (in our case the leaky rectifying linear unit).
To mitigate overfitting, we allow dropout, i.e. probabilistic zeroing of neurons during training.

The relatively small dataset of 1,000 LH simulations makes this a somewhat non-trivial task,
thus we perform automated hyperparameter optimization using the Optuna package~\cite{Akiba2019}
to converge at good architectures.
Generally, our networks have $2 -- 4$ hidden layers with a few hundred neurons each
and high dropout rates $20 -- 70\,\%$.
A useful null-test is the following. From the 1P set we can interpolate the dependence on $\sigma_8$, $\Omega_m$
(simple linear interpolators which will generally be biased due to sample variance).
Using these interpolants, we can remove most of the dependence on the cosmological parameters from
the LH data (c.f. the cyan data points in Fig.~\ref{fig:hmfcorr}).
If the neural net is well-converged, it should matter very little whether the training data have
the full dependence on cosmology or whether it is mostly factored out.
The point here is that we are essentially fitting the same data, just with different transformations
applied.
It is not actually very important what these transformations are as a function of $\sigma_8$ and $\Omega_m$,
we simply chose ones that remove most of the dependence on these cosmological parameters.

Indeed, in the case of SIMBA, this null test is passed satisfactorily.
However, IllustrisTNG is more challenging, since the feedback parameters generally have a much smaller
effect on the $y$ observables in this simulation.
Thus, we need to apply an additional transformation, as described in the following section.

\subsection{Halo model correction factors for IllustrisTNG}
\label{sec:nn:hmf}

\begin{figure*}
\centering
\includegraphics[width=0.8\textwidth]{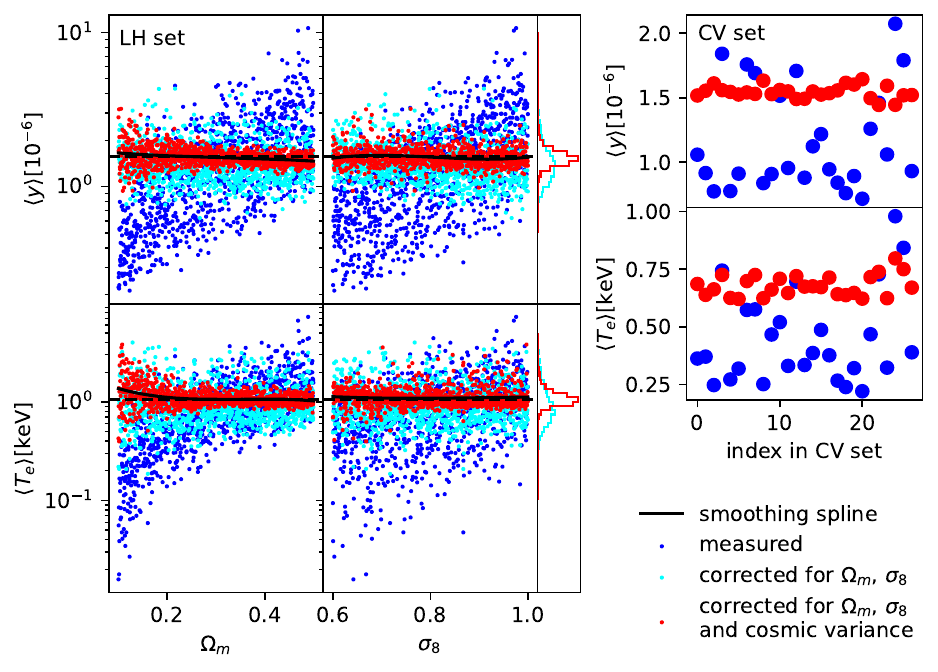}
\caption{Illustration of the effectiveness of analytical correction factors for the IllustrisTNG
         simulation outputs.
         Blue markers are direct measurements from the simulations.
         The cyan markers represent a rescaling to the fiducial cosmology ($\Omega_m=0.3$, $\sigma_8=0.8$),
	 using fits to the CAMELS 1P set.
	 Since the CAMELS CV set is at constant cosmological and astrophysical parameters,
	 the cosmology correction is meaningless there.
	 For the red markers, in addition to the cosmology correction, a correction for sample variance
	 has been applied, as described in the text.
	 The dashed black lines represent the `ground truth' measurements from the larger boxes,
	 while the solid black lines are smoothing splines through the red markers
	 (these are solely to guide the eye).
	 For the LH panels on the left, all data points have been rescaled using Eq.~\eqref{eq:bias}.
	 We have also plotted histograms corresponding to the cyan and red markers.}
\label{fig:hmfcorr}
\end{figure*}

As mentioned before, naively training the neural networks on the IllustrisTNG LH simulations
does not yield robust results according to the null test described above.
The reason is that, in contrast to SIMBA, the dependence on feedback parameters is obscured
by the overwhelming noise due to sample variance.
This issue can be seen in Fig.~\ref{fig:PIXIEcov}, where for IllustrisTNG the 1P data points
scatter much less than the LH points.
There is, however, a simple method to mitigate this problem.
As we have argued in the introduction, the halo model provides a relatively good description
of sparse fields such as Compton-$y$.
Furthermore, the factorization of Eq.~\eqref{eq:factorization} should be a good initial guess.
This motivates the following `correction' formula:
\begin{equation}
\langle y \rangle \leftarrow \langle y \rangle
                             \frac{\langle y \rangle_\text{hm}^\text{Tinker HMF}}
			          {\langle y \rangle_\text{hm}^\text{measured HMF}}\,.
\label{eq:hmfcorr}
\end{equation}
Here, `Tinker HMF' stands for the halo mass function from Ref.~\cite{Tinker2010}
and the expression is analogous for $\langle T_e \rangle$.
On the other hand, the mass function can also be measured in the individual CAMELS simulations.
In order for the numerical integrations to be well behaved, we express it as
\begin{equation}
\frac{dn}{d\log M} \sim \exp\left[-\frac{1}{2}\left(\frac{\log M'}{\sigma}\right)^{\!\!2}\right]
                        \ast \sum_i \delta(\log M' - \log M_i)\,,
\end{equation}
where $\ast$ indicates convolution and we have dropped some prefactors.
Our results are relatively insensitive to the hyperparameter $\sigma$; in the following
we will use $\sigma = 0.19\,\text{dex}$.
The halo masses $M_i$ were measured using a friends-of-friends finder \cite{Davis1985}.
We interpolate the smoothed mass function in mass and redshift using a bilinear routine.

In order to perform the halo model calculation, we use the pressure profile fitting formula
from Ref.~\cite{Battaglia2012} and assume isothermal halos with temperatures according to
Ref.~\cite{Arnaud2005}.
We apply a $20\,\%$ correction to the halo masses entering the temperature fitting formula
to account for hydrostatic mass bias; the same value was used in Ref.~\cite{Hill2015}.
It should be noted that the fitting formula from Ref.~\cite{Arnaud2005} is likely not as
accurate as more recent proposals \citep[e.g.][]{Lee2020},
but as we show in the following, it is sufficient for our purposes.
Following Ref.~\cite{Hill2015}, we assume a radial cut-off at $2.5\,R_\text{vir}$
with the virial radius definition of Ref.~\cite{BryanNorman1998}.

Of course, this procedure rests on a number of assumptions, and we should verify that it
yields reasonable results.
Fig.~\ref{fig:hmfcorr} shows in the right panel in blue the original measurements and
in red the mass function-corrected values for the 27 realizations in the CV set.
Indeed, we see that the corrected values scatter much more tightly;
the fact that the procedure is more successful for $\langle y \rangle$ than $\langle T_e \rangle$
is likely because the temperature fitting formula used is calibrated at larger halo
masses than those that dominate the signal.
In the left panel of Fig.~\ref{fig:hmfcorr}, we show projections of the LH set onto the $\sigma_8$
and $\Omega_m$ axes.
As expected, the blue measurement data points have a large scatter and show a mean
evolution with the cosmological parameters.
The cyan data points illustrate our procedure for dividing out most of the
cosmology dependence, as described in the previous section.
However, they still display a large scatter which is dominated by sample variance.
Finally, the red data points have the mass function correction applied in addition.
It is striking how much tighter they cluster compared to the cyan markers, again indicating
that the mass function correction is working rather well.
The scatter is relatively independent of cosmology, except for the low-$\Omega_m$ points
where the halo model approach seems to start to break down.
We observe that the remaining scatter in the LH panels significantly exceeds what is observed
for the red markers in the CV panels, demonstrating that applying the mass function correction
yields an LH set that is dominated by signal from the feedback parameters.
Indeed, the neural networks trained on these de-noised LH simulations
pass the null test described in Sec.~\ref{sec:nn:training}.

It is worth noting that we tried the described mass function correction procedure for SIMBA
as well.
However, for these simulations it did not work well at all, indicating that halos in SIMBA
are not well approximated by the fitting formulae we used in the halo model
(this may be related to the very long-range effects feedback has in SIMBA \cite{Borrow2020}).
Fortunately, the signal from feedback parameters is strong enough in SIMBA that it was possible
to extract from the LH data with full sample variance contamination.

\section{Results and Discussion}
\label{sec:results}

In the previous section, we have constructed interpolating neural networks which return
the $y$ distortion monopoles $\langle y \rangle$ and $\langle T_e \rangle$ as functions
of astrophysical feedback parameters $A_\text{SN1}$, $A_\text{AGN1}$, $A_\text{SN2}$, $A_\text{AGN2}$
and cosmological parameters $\Omega_m$, $\sigma_8$.
We now aim to explore the dependence on the astrophysical parameters and translate it into
forecast constraints assuming the spectral distortion measurement described in Sec.~\ref{sec:PIXIE}.

\subsection{Parameter dependence}
\label{sec:results:pardep}

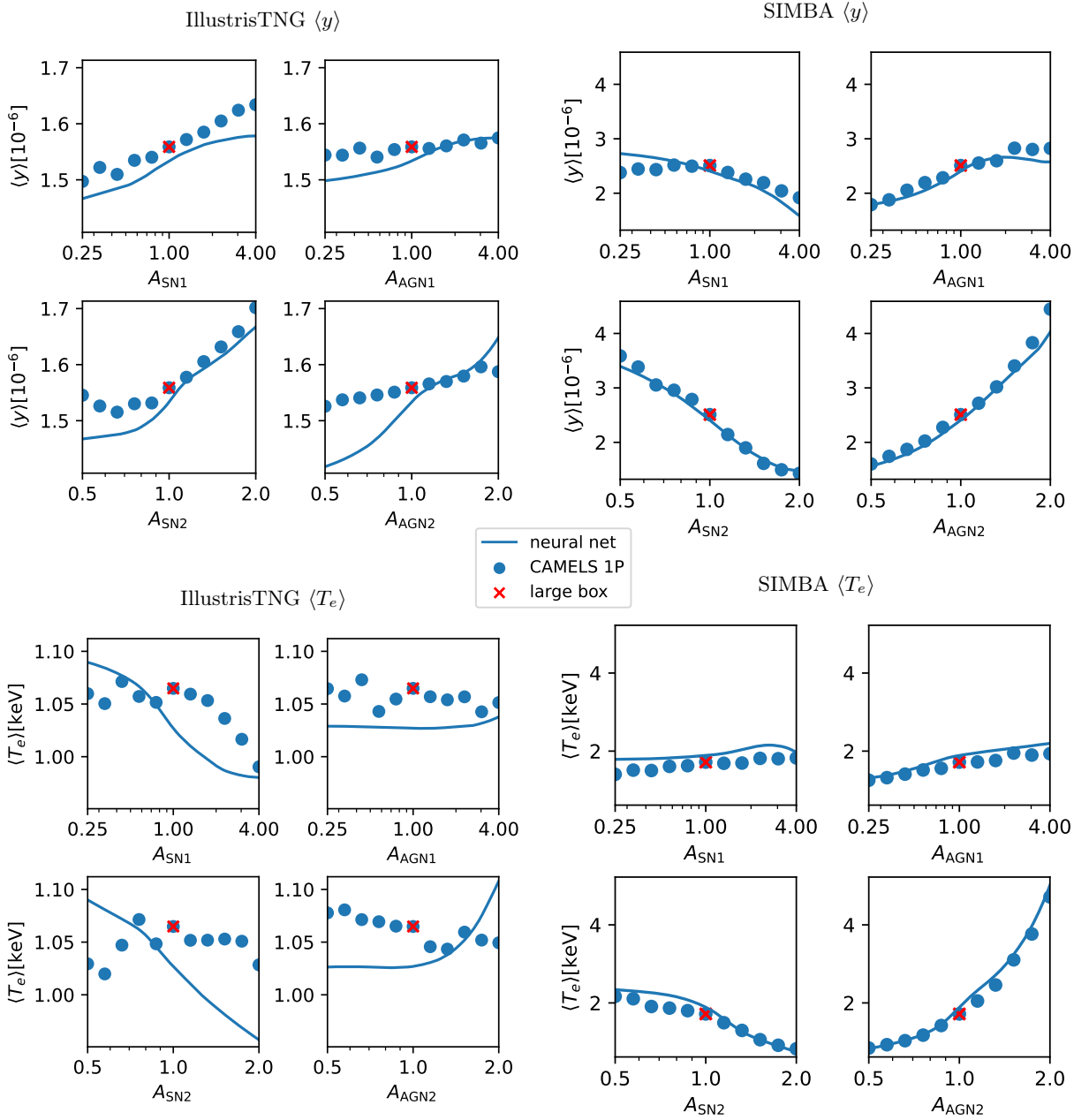
\begin{figure*}
\centering
\begin{tikzpicture}
	\draw (0, 0) node (yIllustrisTNG) [above left] {\includegraphics[width=0.45\textwidth]{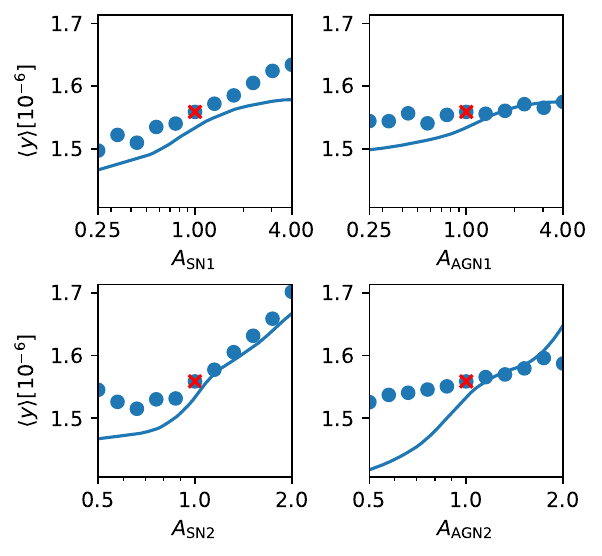}};
	\draw (yIllustrisTNG.east) node (ySIMBA) [right] {\includegraphics[width=0.45\textwidth]{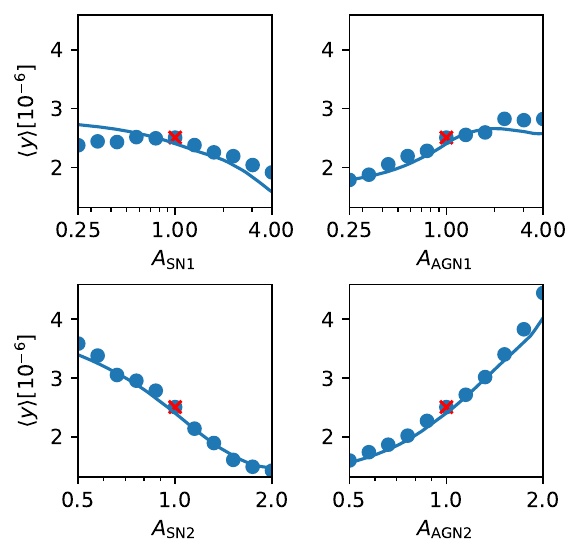}};

	\draw (0, 0) node (legend) [yshift=-0.25cm,xshift=0.2cm] {\includegraphics[width=0.15\textwidth]{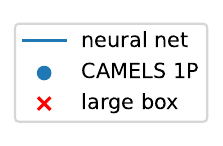}};

	\draw (yIllustrisTNG.south) node (TIllustrisTNG) [below,yshift=-1cm] {\includegraphics[width=0.45\textwidth]{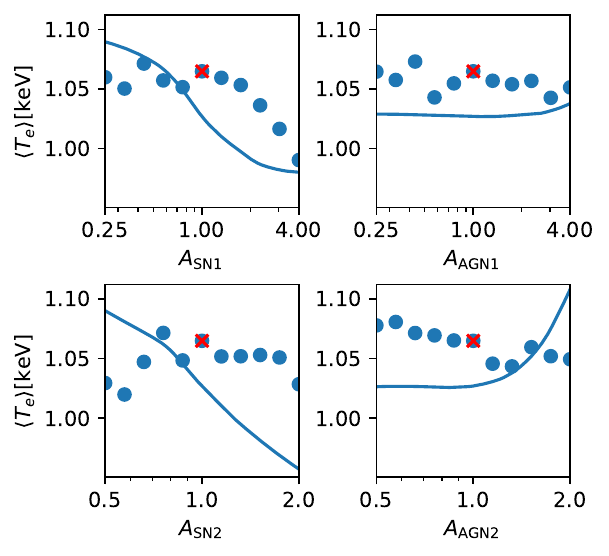}};
	\draw (TIllustrisTNG.east) node (TSIMBA) [right] {\includegraphics[width=0.45\textwidth]{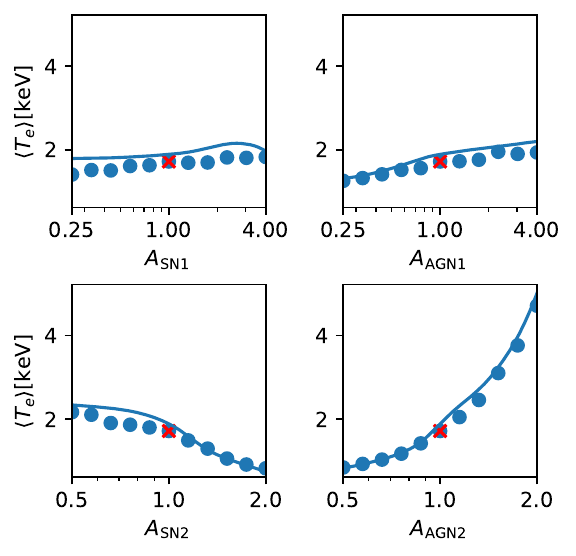}};
	
	\draw (yIllustrisTNG.north) node [above] {IllustrisTNG $\langle y \rangle$};
	\draw (TIllustrisTNG.north) node [above] {IllustrisTNG $\langle T_e \rangle$};
	\draw (ySIMBA.north) node [above] {SIMBA $\langle y \rangle$};
	\draw (TSIMBA.north) node [above] {SIMBA $\langle T_e \rangle$};

\end{tikzpicture}
\caption{Dependence of spectral distortions on individual simulation feedback parameters.
         In each panel, all cosmological and astrophysical parameters are held fixed
	 apart from the one shown on the horizontal axis.
         The blue markers are measurements from the CAMELS 1P set, rescaled such that at
	 the fiducial model they agree with the measurements from the larger boxes,
	 which are indicated by the red crosses.
	 The blue lines are evaluations of neural networks trained on the CAMELS LH set.
	 We argue in the text that the LH-trained networks are likely more robust than
	 the 1P data in most cases.
	 Note that the vertical plot ranges are identical within each quadrant.}
\label{fig:1P}
\end{figure*}

We first explore how $\langle y\rangle$ and $\langle T_e\rangle$ depend on individual
feedback parameters with the remaining ones (as well as $\sigma_8$ and $\Omega_m$) fixed.
In Fig.~\ref{fig:1P}, we show as blue markers the measurements from the CAMELS 1P set, rescaled such that, their value at the
fiducial point, is equal to that from the corresponding large box (indicated by the red crosses).
The blue lines in Fig.~\ref{fig:1P} are evaluations of the trained neural nets.
Agreement between the 1P results and the neural nets is generally better for SIMBA,
because the observables depend much more strongly on the feedback parameters than in IllustrisTNG.
For IllustrisTNG, there are substantial differences, although in many cases the qualitative trends
are similar in 1P and the neural nets.
As we have argued before, we believe that generally the neural nets are more robust since they marginalize
over the simulation initial conditions.
However, this may not be true for some of the cases in IllustrisTNG for which the dependence is
comparatively weak, because the neural net may have little incentive to accurately capture this
dependence.
There are, however, strong counterarguments against this hypothesis.
It seems clear from the SIMBA panels that in almost all cases the SN and AGN parameter pairs each work
in the same direction.
This behaviour is mirrored in IllustrisTNG, also in the cases where strong deviations from the 1P
set are observed.
We emphasize that our neural net architecture has no implicit preference for grouping the parameters
in such a way.
Furthermore, we have evidence that the initial conditions for the 1P set happen to be somewhat
anomalous. Visually, this manifests itself by the presence of an unusally large void in the $z=0$
density field. More quantitatively, this can be seen in Fig.~\ref{fig:PIXIEcov}, where the 1P data
points (for both IllustrisTNG and SIMBA) tend to fall at higher $\langle T_e \rangle$ than the
mean trend in the LH set.
Finally, as mentioned in Sec.~\ref{sec:nn:training}, we have performed a non-trivial null-test on
the trained networks which is passed satisfactorily
(after correcting for sample variance in the case of IllustrisTNG).
Thus, we believe that in fact for all cases the neural nets are closer to reality than the 1P set.

Turning now to physical interpretation, we observe striking differences between SIMBA and IllustrisTNG.
In the case of $\langle y \rangle$, IllustrisTNG has positive slopes for all four feedback parameters,
while for SIMBA the SN parameters exhibit negative slopes.
It seems rather natural that slopes should generally be positive because integrated electron pressure
is a measure of thermal energy.
However, it has also been established that stronger supernova feedback can counteract the AGN contribution,
since the outflow of matter due to stellar feedback limits the black hole growth rate
\cite{DAA2017} (see also \cite{Booth2013}).
Ref.~\cite{Booth2013} used the OWLS subgrid model which in terms of $\langle y \rangle$ and $\langle T_e \rangle$
is closer to SIMBA than to IllustrisTNG.
This is consistent with the frequency of negative slopes with $A_{\text{SN}i}$ in Fig.~\ref{fig:1P}.

Interestingly, in some cases there are saddle points.
That these are probably mostly real and not artifacts in the neural networks is supported by the fact
that they are also visible in the 1P markers.
Fortunately, these saddle points are generally away from the fiducial model
(except for the AGN dependence of $\langle T_e \rangle$ in IllustrisTNG),
so they will not contaminate the Fisher forecasts significantly.

As we show in Appendix~\ref{app:pairs}, the leading-order behaviour
of the $y$-distortion observables as a function of the four feedback parameters
is well-described by a factorization,
\begin{equation}
x_i = \prod_j g_i^{(j)}(A_j)\,,
\label{eq:separable}
\end{equation}
where the $g_i^{(j)}$ are identical to the 1-parameter functions in Fig.~\ref{fig:1P} up to normalization. 

As a cross-check, we also test our results using another machine learning tool called symbolic regression.
The results, presented in Appendix~\ref{app:sr} support the factorization approximation Eq.~\eqref{eq:separable}.

\subsection{Fisher forecasts}
\label{sec:results:fisher}

We now turn to forecasting constraints on the simulation feedback parameters given our fiducial
model for the PIXIE constraints on $\langle y \rangle$, $\langle T_e \rangle$ presented in
Sec.~\ref{sec:PIXIE}.
It should be emphasized that the primary goal of this exercise is to demonstrate the substantial
constraining power of a realistic $y$-distortion measurement on our understanding of astrophysical
feedback processes.
In particular, the uncertainty on the fiducial model 
implies that the constraints presented in the following should only be interpreted as qualitative
indicators.
We make the approximation that the cosmological model (specifically $\Omega_m$ and $\sigma_8$)
is known perfectly;
in comparison to our ignorance on the subgrid model this is a good assumption.

As we have already seen before, the feedback parameters have very different consequences in
IllustrisTNG and SIMBA.
We will need to assume that the chosen priors on these parameters are reasonable in terms
of other observables, so that a direct comparison between the constraints
(in units of the prior) makes any sense at all.

First, we construct the $4 \times 4$ Fisher matrices for IllustrisTNG and SIMBA as usual,
\begin{equation}
F_{ab} = \frac{\partial x^i}{\partial A^a}
C^{-1}_{ij}\,
\frac{\partial x^j}{\partial A^b}\,,
\end{equation}
with $C$ the $2 \times 2$ covariance matrix, marginalized over all foreground nuisance parameters
(c.f. Sec.~\ref{sec:PIXIE}), and the derivatives are computed from the neural network interpolators
with finite difference step sizes chosen to match the scale of the posterior.
Of course, these Fisher matrices are degenerate since two measurements are not sufficient
to constrain four parameters.

For each simulation type, by diagonalizing the Fisher matrix we identify
the best-constrained orthogonal parameter combinations as
\begin{equation}
\begin{aligned}
A^{(1)}_\text{IllustrisTNG}& = A_\text{SN1}^{+0.20} A_\text{AGN1}^{+0.13}
                               A_\text{SN2}^{+0.77} A_\text{AGN2}^{+0.59}\,, \\
A^{(2)}_\text{IllustrisTNG}& = A_\text{SN1}^{+0.52} A_\text{AGN1}^{-0.10}
                               A_\text{SN2}^{+0.44} A_\text{AGN2}^{-0.73}\,, \\
A^{(1)}_\text{SIMBA}& = A_\text{SN1}^{-0.17} A_\text{AGN1}^{+0.30}
                        A_\text{SN2}^{-0.72} A_\text{AGN2}^{+0.61}\,, \\
A^{(2)}_\text{SIMBA}& = A_\text{SN1}^{+0.42} A_\text{AGN1}^{-0.37}
                        A_\text{SN2}^{+0.36} A_\text{AGN2}^{+0.74}\,.
\end{aligned}
\label{eq:paramcomb}
\end{equation}
Note that our methodology of finding the most constrained parameter combination(s)
is similar to that used in galaxy photometric survey studies for identifying the
$S_8 \equiv \sigma_8 \Omega_m^{\alpha}$ parameter combination \cite{Amo21,Jai97}.
The induced $2 \times 2$ Fisher matrix on the thus identified subspaces tangent to the
fiducial model is not singular.
Of course, these combinations are not unique since the Fisher matrix only gives linear-order information.
Not surprisingly given the intuition from the upper half of Fig.~\ref{fig:1P},
in $A^{(1)}$ the powers have identical signs for IllustrisTNG,
while for SIMBA the SN and AGN parameters have powers of opposite signs.

We compute the $68\,\%$ constraints
\begin{align}
\sigma(A^{(1)}_\text{IllustrisTNG})& = 0.015\,, &\sigma(A^{(1)}_\text{SIMBA})& = 0.0024\,, \nonumber\\
\sigma(A^{(2)}_\text{IllustrisTNG})& = 1.3\,,   &\sigma(A^{(2)}_\text{SIMBA})& = 0.075\,. \nonumber
\end{align}
Thus, a PIXIE-like experiment could place percent-level constraints on parameter combinations
that are currently only known to little better than an order of magnitude.
In the case of SIMBA the measurement of the relativistic component would also allow a $\sim 10\,\%$
constraint on a second parameter combination.
This is not possible for IllustrisTNG, due to the extremely small variation of $\langle T_e \rangle$.
Conversely, this implies that a measurement of $\langle T_e \rangle$
significantly different from the fiducial model
has the potential to simply rule out the CAMELS-based implementation of the IllustrisTNG model.

\subsubsection{Robustness check}

We have argued before (Sec.~\ref{sec:results:pardep} that we believe the neural network interpolators
to be more robust than the 1P data, particularly for IllustrisTNG where they disagree substantially.
As a robustness check, we have repeated the Fisher analysis using derivatives estimated from
one-dimensional interpolators through the 1P set.
In the case of IllustrisTNG, the error bar on the best-constrained parameter combination inflates
by a factor $\sim 3$, while the orthogonal combination remains unconstrained.
The slight inflation is due to the derivatives from the 1P set being slightly shallower
than those from the neural net, see Fig.~\ref{fig:1P}.
In the case of SIMBA, the error bars increase by about $20\,\%$.
The corresponding parameter combinations differ from the ones listed in Eq.~\eqref{eq:paramcomb},
but the qualitative features (relative magnitudes and signs of the exponents) are very similar.
These comparisons indicate that our results for IllustrisTNG are robust within a factor of a few
while those for SIMBA are very accurate.

\section{Conclusions}
\label{sec:concl}

This work is one of the first systematic studies of the information content in the
late-time Sunyaev-Zel'dovich spectral distortions.
Besides a simulation-based forecast in the context of specific subgrid models,
in the appendices we have also given some novel theoretical results regarding
the IGM and reionization contributions.

We have measured the non-relativistic and relativistic mean spectral distortion amplitudes
$\langle y \rangle$ and $\langle T_e \rangle$ in the CAMELS simulations suite.
By training neural networks on this data, we have constructed interpolators returning the
two signals as a function of $\Omega_m$, $\sigma_8$, and four feedback parameters.
In the case of IllustrisTNG, the observables depend only weakly on feedback, necessitating
the use of halo model-derived correction factors to reduce the large sample variance
due to the small CAMELS box size.

Incidentally, the described method to reduce scatter arising from small simulation boxes
by using mass function-dependent scaling factors
should be more generally applicable to many works concerned with fields that can be approximated
with a halo model.
We have also tried to reduce biases as much as possible by matching our data to the
comparatively much larger size flagship IllustrisTNG and SIMBA simulations.

Using the interpolating neural networks, we have performed a Fisher forecast assuming a
Gaussian posterior on $\langle y \rangle$ and $\langle T_e \rangle$, which had been computed
assuming a PIXIE-like experiment and a realistic foreground model.
Of course, our work is relevant for experiments other than PIXIE, e.g.\ ESA's Voyage 2050
large-scale proposal (see e.g.~\cite{Chluba2021}),
which is expected to reach even tighter constraints on distortion parameters.

We find that in the case of IllustrisTNG, only a single parameter combination can be strongly constrained,
at the $\sim 2\,\%$ level.
On the other hand, in the case of SIMBA,
the availability of two data points enables two orthogonal combinations to be measured,
to $\sim 0.2\,\%$ and $\sim 8\,\%$.
We emphasize again that although the feedback parameters are qualitatively similar between
IllustrisTNG and SIMBA, their detailed meaning differs substantially so direct comparisons
must be carried out with caution.

Given the limited information content from only two observables,
it is interesting to ask what other measurements could be added in order to improve
the constraints presented here.
The tSZ and kSZ profiles of halos are already being used in order to assess the viability
of simulation models, and Ref.~\cite{Moser2022} demonstrates using CAMELS that next-generation
CMB experiments could place constraints on the feedback parameters.
Similarly, Ref.~\cite{Wadekar2022} shows that deviations from self-similarity in the integrated tSZ flux $-$ halo mass relation ($Y_\textup{SZ}-M$)
can also be used to constrain feedback parameters.
A combination of such constraints could substantially improve upon the error bars presented in this work.
Furthermore, the CAMELS feedback parameters also affect observables beyond the SZ effects,
like the properties of galaxies.
Typically observations of such astrophysical nature are difficult to propagate into hard posteriors,
but at least qualitatively the check for simultaneous viability of a given subgrid model should
be very useful for simulators.

The main source of uncertainty in the Fisher forecast is likely due to errors in the interpolators.
We have performed extensive consistency checks on the trained neural nets, but the limited
number of data points in the 1,000 CAMELS LH simulations places fundamental limits in the
possible accuracy.
For this reason, the given constraints have some associated uncertainty,
of order unity for IllustrisTNG and at the $10\,\%$ level for SIMBA.
Nonetheless, our results are indicative of the transformative effect that measurements of the low-redshift
spectral distortions could have for our understanding of baryonic feedback.

\acknowledgments{
LT thanks Will Coulton, Shy Genel, and David Spergel for useful discussions.
DW gratefully acknowledges support from the Friends of the Institute for Advanced Study Membership.
The Flatiron Institute is supported by the Simons Foundation.
JCH acknowledges support from NSF grant AST-2108536.
DAA was supported in part by NSF grants AST-2009687 and AST-2108944.
JC was supported by the Royal Society as a University Research fellow (No.~URF/R/191023) and by the ERC Consolidator Grant {\it CMBSPEC} (No.~725456).
NB acknowledges support from NSF grant AST-1910021 and NASA grants
21-ADAP21-0114 and 21-ATP21-0129.
}

\appendix

\section{Electron temperature and the relativistic SZ effect}
\label{app:rSZ-weighting}
A cluster's thermal SZ contribution for a given line of sight can be written as 
\begin{equation}
\Delta I_\nu(\hat{n}) = I_0 \int N_e(\hat{n}, l)\, \sigma_{\rm T} \, S_\nu\left(T_e(\hat{n}, l)\right)\,dl
\end{equation}
where $I_0=B_\nu(T_0)$ is the CMB blackbody spectrum, $N_e$ denotes the electron number density, $dl$ parameterizes the line of sight integration and $S_\nu(T)$ determines the SZ spectrum with relativistic temperature corrections. If the electron temperature is constant along the line of sight, $T_e(\hat{n}, l)\equiv T_e(\hat{n})$, one can simply write $\Delta I_\nu(\hat{n})=\tau(\hat{n}) I_0
S_\nu\left(T_e(\hat{n})\right)$, where the Thomson optical depth is $\tau(\hat{n})=\int N_e(\hat{n}, l)\, \sigma_{\rm T} \,dl$. However, generally the temperature varies along the line of sight, such that a moment expansion provides a simpler method for analyzing and describing the SZ signal \citep{Chluba2012moments}. 

Defining $S_\nu^{(k)}(T)\equiv \partial^k S_\nu(T)/\partial T^k$, we can perform the moment expansion of the SZ signal around the $\tau$-weighted temperature, $T_e^{\tau}(\hat{n})=\int N_e(\hat{n}, l)\, \sigma_{\rm T} T_e(\hat{n}, l)\,dl/\tau(\hat{n})$. Up to second order in temperature,  this yields
\begin{equation}
\Delta I_\nu(\hat{n}) \approx \tau I_0 \left\{S_\nu\left(T^\tau_e\right)
+\frac{1}{2} S_\nu^{(2)}\left(T^\tau_e\right)
\left[\left<T^2_e\right>_{\tau}-(T^\tau_e)^2\right]\right\},
\end{equation}
where we suppressed the dependence on $\hat{n}$ and introduced the $\tau$-weighted temperature moments
\begin{equation}
\left<T^k_e\right>_{\tau}=\frac{\int N_e(\hat{n}, l)\, \sigma_{\rm T} T^k_e(\hat{n}, l)\,dl}{\tau(\hat{n})}
\end{equation}
such that $T_e^{\tau}\equiv \left<T_e\right>_{\tau}$. For the standard $\Lambda$CDM cosmology the first two temperature moments from halos alone are $\left<T_e\right>_{\tau}=0.208\, {\rm keV}$ and $\left<T^2_e\right>_{\tau}=0.299\, {\rm keV}^2$ \citep{Hill2015}.

Alternatively, we can perform the moment expansion around the $y$-weighted temperature, $T_e^{y}$. By introducing the $y$-weighted moments
\begin{equation}
\left<T^k_e\right>_{y}=\frac{\int N_e(\hat{n}, l)\, \sigma_{\rm T} \frac{T_e(\hat{n}, l)}{m_e}\, T^k_e(\hat{n}, l)\,dl}{y(\hat{n})}
\end{equation}
with $y(\hat{n})=\int N_e(\hat{n}, l)\, \sigma_{\rm T} \frac{T_e(\hat{n}, l)}{m_e}\,dl$ and $T_e^{y}\equiv \left<T_e\right>_{y}$, again to second order in the temperature one then has
\begin{equation}
\Delta I_\nu(\hat{n}) \approx y I_0 \left\{\tilde{S}_\nu\left(T^y_e\right)
+\frac{1}{2} \tilde{S}_\nu^{(2)}\left(T^y_e\right)
\left[\left<T^2_e\right>_{y}-(T^y_e)^2\right]\right\}.
\end{equation}
where $\tilde{S}_\nu=S_\nu/\Theta_e$, $\tilde{S}_\nu^{(k)}=\partial^k \tilde{S}_\nu(T)/\partial T^k$,
and we introduced the dimensionless temperature $\Theta_e=\frac{T_e}{m_e}$.
These two representations are essentially equivalent\footnote{They become indistinguishable when more temperature terms are included in the expansion.}; however, the latter is slightly more economic when it comes to capturing the relativistic tSZ effect, as we will see next. 

\textit{Low temperature limit and mix of hot and cold gas:}
In the low temperature limit one can write \citep[e.g.,][]{Sazonov1998, Itoh1998}
\begin{equation}
S_\nu\left(T\right) \approx \Theta_e Y_0(x) + \Theta^2_e Y_1(x)
\end{equation}
where $x=\frac{h\nu}{T_0}$ and the functions $Y_0$ and $Y_1$ are the first two terms of the asymptotic expansion for the tSZ signal \citep[e.g., see][]{Itoh1998}. It is clear that in this limit, only two independent spectral parameters can be determined. This is directly evident when using the $y$-weighted moments with $\tilde{S}^{(2)}_\nu\left(T\right)\approx 0$:
\begin{equation}
\frac{\Delta I_\nu(\hat{n})}{I_0} \approx y S_\nu\left(T^y_e\right)
=y \left[Y_0(x)+ \frac{T_e^{y}}{m_e} Y_1(x)\right].
\end{equation}
Since $T^y_e\equiv\left<T^2_e\right>_{\tau}/\left<T_e\right>_{\tau}$, the two important parameters for the spectral analysis in the $\Lambda$CDM case are expected to be $y\approx 1.77\times 10^{-6}$ and $T^y_e\approx 1.44\, {\rm keV}$~\cite{Hill2015}. However, the presence of extremely cold gas from reionization modifies the observational inference. This can be seen by adding another $y$-distortion contribution with no relativistic correction
\begin{equation}
\frac{\Delta I^{\rm tot}_\nu(\hat{n})}{I_0} \approx y^{\rm re}\,Y_0(x)+y \left[Y_0(x)+ \frac{T_e^{y}}{m_e} Y_1(x)\right].
\end{equation}
The reionization $y$-parameter, $y^{\rm re}$, is roughly $10\%$ of the cluster contribution \citep{Hu1994pert, Hill2015} and cannot be distinguished from the cluster contribution with distortion measurements alone. 
In an analysis, an effective electron temperature of $T_e^{y, *}=T_e^{y}/(1+y^{\rm re}/y)\approx 1.30\, {\rm keV}$ would thus be recovered. Due to small contributions from higher order temperature corrections, the recovered result for $\Lambda$CDM is $T_e^{y, *}\approx 1.24\, {\rm keV}$ \citep{Abitbol2017}.

\section{Cosmology dependence}
\label{app:pwrlaws}

Using the halo model we compute how the distortion monopoles $\langle y \rangle$ and $\langle T_e \rangle$
depend on the relevant $\Lambda$CDM parameters.
We use the halo model fitting formulae described in Sec.~\ref{sec:nn:hmf} to compute the observables
$x_i=\{\langle y \rangle, \langle T_e \rangle\}$ as functions of
$\theta_j=\{h, \Omega_m, \Omega_b, n_s, \sigma_8\}$.
The resulting functions are well approximated by power laws,
$x_i = x_i^{(0)} (\theta_j/\theta_j^{(0)})^{\alpha_{ij}}$.
The resulting one-parameter fits are given by:
\begin{equation}
\begin{aligned}
10^6 \langle y \rangle(h)& = 1.48 (h/0.6711)^{1.59}\,, \\
\text{keV}^{-1}\langle T_e \rangle(h) & = 1.51 (h/0.6711)^{-0.86}\,, \\
10^6 \langle y \rangle(\Omega_m)& = 1.48 (\Omega_m/0.3)^{0.92}\,, \\
\text{keV}^{-1}\langle T_e \rangle(\Omega_m) & = 1.51 (\Omega_m/0.3)^{0.36}\,, \\
10^6 \langle y \rangle(\Omega_b)& = 1.48 (\Omega_b/0.0490)^{0.84}\,, \\
\text{keV}^{-1}\langle T_e \rangle(\Omega_b) & = 1.51 (\Omega_b/0.0490)^{0.09}\,, \\
10^6 \langle y \rangle(n_s)& = 1.48 (n_s/0.9624)^{1.52}\,, \\
\text{keV}^{-1}\langle T_e \rangle(n_s) & = 1.51 (n_s/0.9624)^{-0.84}\,, \\
10^6 \langle y \rangle(\sigma_8)& = 1.48 (\sigma_8/0.8)^{3.86}\,, \\
\text{keV}^{-1}\langle T_e \rangle(\sigma_8) & = 1.51 (\sigma_8/0.8)^{1.93}\,.
\end{aligned}
\end{equation}
Note that the given scalings have a number of uncertainties.
First, they only include the halo contribution.
Second, the assumed cluster temperature model may not be optimal, as discussed in Sec.~\ref{sec:nn:hmf}.
Thus, we caution against blind use of these equations.
However, for small deviations from the pivot cosmology they should provide
reasonable approximations for the slope, even though the prefactors are most likely
not very useful.

\section{Reionization contribution}
\label{app:reio}

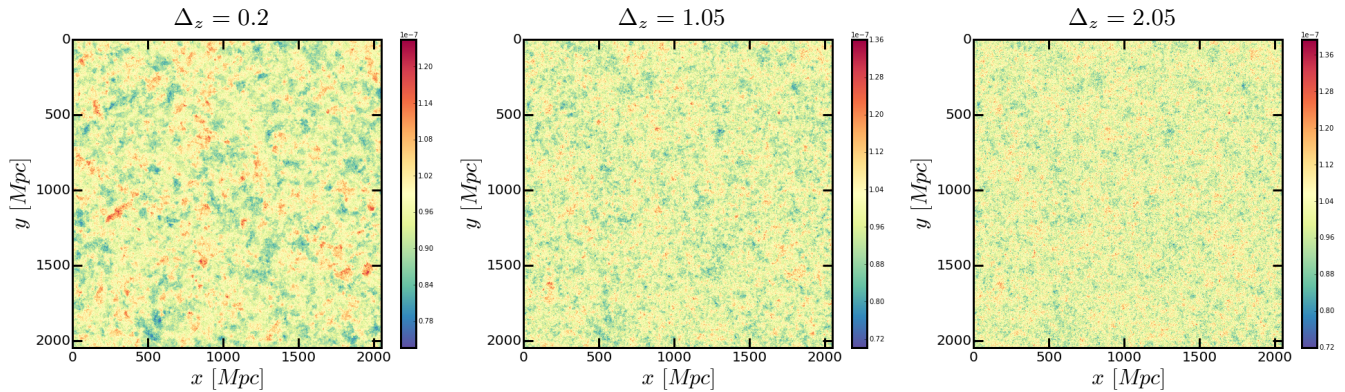
\begin{figure*}
\centering
\begin{tikzpicture}
\draw (0, 0) node (fidduration)
      {\includegraphics[width=0.32\textwidth]{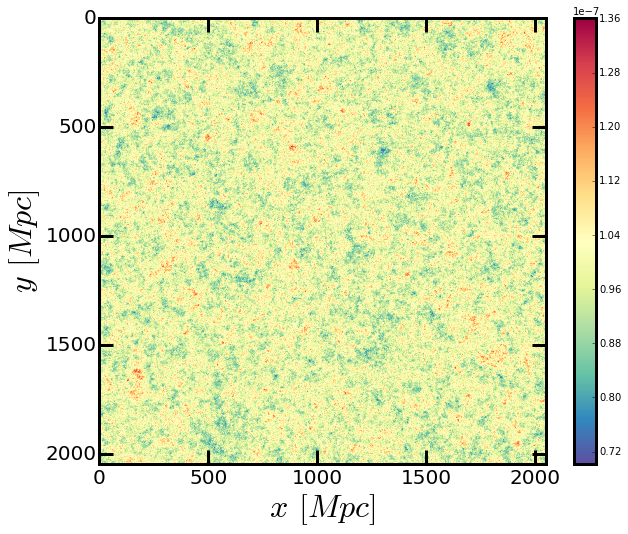}};
\draw (fidduration.west) node [left] (shortduration) 
      {\includegraphics[width=0.32\textwidth]{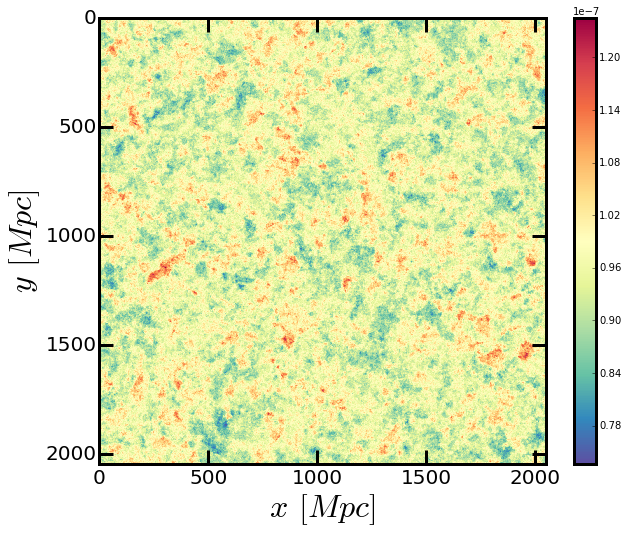}};
\draw (fidduration.east) node [right] (longduration) 
      {\includegraphics[width=0.32\textwidth]{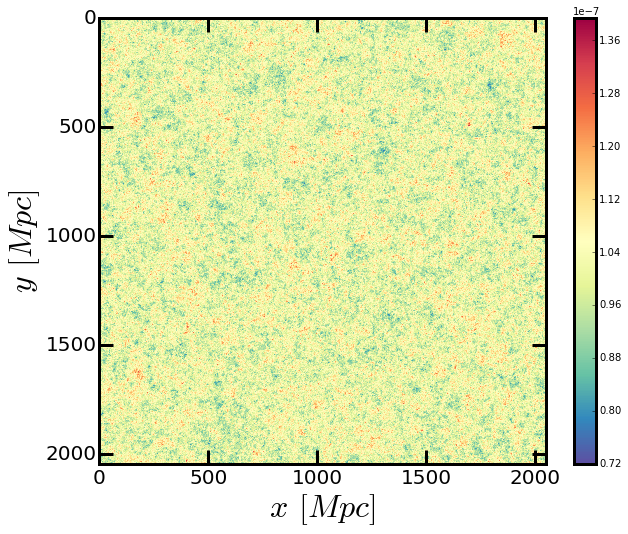}};
\draw (fidduration.north) node {$\Delta_z = 1.05$};
\draw (shortduration.north) node {$\Delta_z = 0.2$};
\draw (longduration.north) node {$\Delta_z = 2.05$};
\end{tikzpicture}
\caption{Maps of the Compton-$y$ field from reionization, generated using a semi-analytic model
         on a gravity-only simulation.
         The panels assume different durations of reionization, with the short, fiducial, and long
	 duration models from left to right.
	 Note the slightly different color scales.}
\label{fig:reio}
\end{figure*}

The non-relativistic $\langle y \rangle$ receives a small contribution from the epoch of reionization,
which is not modelled in our simulations.
This appendix describes an estimate of the magnitude and uncertainty of this effect.
We calculate the $\langle y \rangle$ from reionization using maps constructed by ray-tracing through
the past light cone of a semi-analytic realization from $z=5.5$ to $z=20$, which defines the redshift
range we consider for reionization.
In this semi-analytic realization, reionization fields are constructed on a gravity-only simulation
following the method in Ref.~\cite{Battaglia2013}.
For each spatial cell in the $N$-body simulation we have a density and ionization state as a function
of time.
We set the initial temperature of the cells when they reionize as $T_0=2 \times 10^4\,\text{K}$,
then the $i$th cell cools adiabatically according to
\begin{equation}
\frac{T_i(z)}{T_0} = \frac{1+z}{1+z_{i,\text{RE}}}\,,
\end{equation}
where $z_{i,\text{RE}}$ is the redshift at which the given cell reionizes.
Using the parametric reionization model from Ref.~\cite{Battaglia2013},
we change the duration and midpoint of reionization from the fiducial parameters.
We show Compton-$y$ maps from reionization for our fiducial model and the two extreme duration models,
short and long duration, in Fig.~\ref{fig:reio}.
For the fiducial reionization model,
with a median reionization redshift of $z_\text{mid}=10$, where $50\,\%$ of the Universe has reionized by mass,
and a duration parameter $\Delta_z=1.05$,
we find that $\langle y \rangle_\text{reio}=9.9 \times 10^{-8}$.
For the short ($\Delta_z=0.2$) and long ($\Delta_z=2.05$) duration models
we find $9.6 \times 10^{-8}$ and $1.0 \times 10^{-7}$, respectively,
while the dependence on $z_\text{mid}$ (which we varied in $[8, 12]$) is smaller.
Thus, reionization contributes less than $10\,\%$ to the total signal and we estimate the uncertainty
as $\sim 5 \times 10^{-9}$, significantly below the error budget for our assumed experiment.

Clearly, the relativistic distortion receives miniscule contributions from reionization
as $T_e/m_e \lesssim 10^{-4}$.

\section{IGM contribution}
\label{app:IGM}

In this appendix we compute the contribution from the intergalactic medium (IGM) to the $\langle y \rangle$
signal.
We define this quantity as all Compton-$y$ generated at $z<5.5$ (which marks the end of reionization,
c.f. Appendix~\ref{app:reio}) outside of any halo.
Since the IGM pressure is low compared to the ICM, a reasonable approximation is
\begin{equation}
\langle y \rangle_\text{IGM} = \frac{\sigma_\text{T}}{m_e} \int dl\,\bar{n}_e(z) T_\text{IGM}(z)
\end{equation}
where the integration is over distance up to $z=5.5$,
$\bar{n}_e(z) \equiv x_e(z) \Omega_b \rho_\text{crit}(z)$,
and $x_e(z)$ is the free electron fraction.
In our fiducial model, we assume $T_{\text{IGM},0}=2 \times 10^4\,\text{K}$ for the IGM temperature
at the end of reionization.
We then assume the temperature drops adiabatically.
The function $x_e(z)$ will depend on the redshift $z_\text{HeII}$ at which HeII reionizes.
In our fiducial model we assume instantaneous HeII reionization at $z_\text{HeII}=3.5$ such that
\begin{equation}
x_e(z) = \begin{cases}
	 \frac{3X_H+1}{4}; &z > z_\text{HeII}\,, \\
	 \frac{X_H+1}{2};  &z < z_\text{HeII}\,,
         \end{cases}
\end{equation}
where $X_H$ is the primordial hydrogen mass fraction.
The fiducial model yields $\langle y \rangle_\text{reio}=7.1 \times 10^{-8}$.
Changes in the starting value of $T_\text{IGM}$ linearly affect the Compton-$y$.
Assuming $z_\text{HeII}=2.5$ results in $7 \times 10^{-8}$
and $z_\text{HeII}=4.5$ gives $7.2 \times 10^{-8}$.
Thus, the IGM contribution is in magnitude comparable to the reionization signal
with somewhat larger theoretical uncertainty driven by the temperature normalization.

\section{Two-parameter dependence}
\label{app:pairs}

\begin{figure*}
\centering
\begin{tikzpicture}
	\draw (0, 0) node (yIllustrisTNG) {\includegraphics[width=0.45\textwidth]{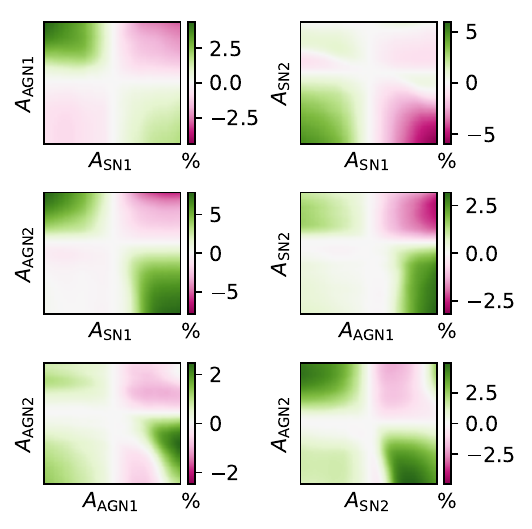}};
	\draw (yIllustrisTNG.south) node (TIllustrisTNG) [below,yshift=-1cm] {\includegraphics[width=0.45\textwidth]{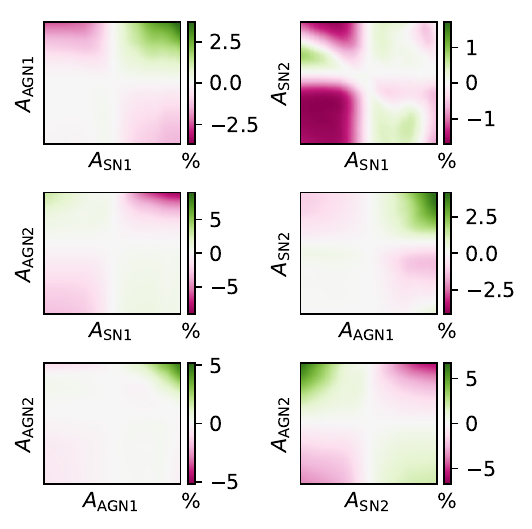}};
	\draw (yIllustrisTNG.east) node (ySIMBA) [right] {\includegraphics[width=0.45\textwidth]{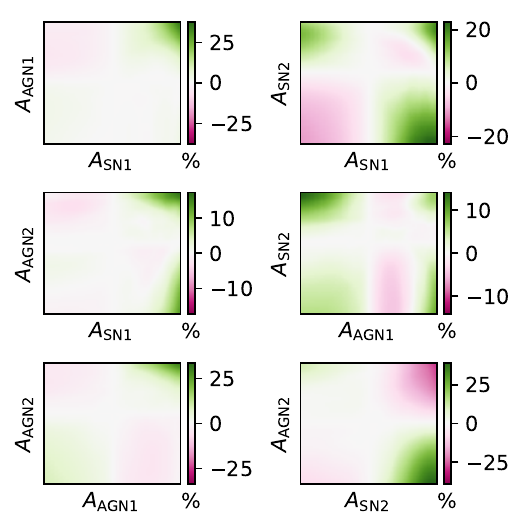}};
	\draw (TIllustrisTNG.east) node (TSIMBA) [right] {\includegraphics[width=0.45\textwidth]{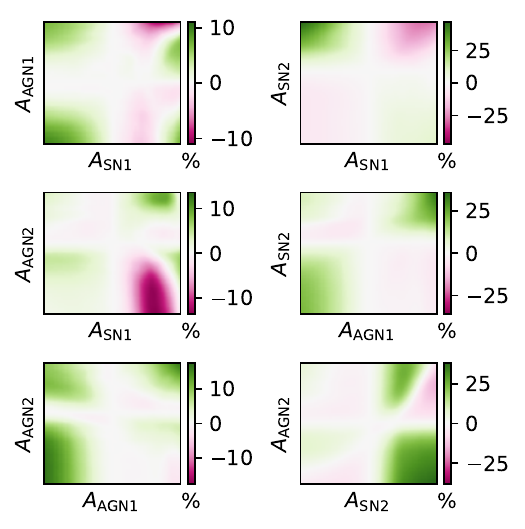}};
	
	\draw (yIllustrisTNG.north) node [above] {IllustrisTNG $\langle y \rangle$};
	\draw (TIllustrisTNG.north) node [above] {IllustrisTNG $\langle T_e \rangle$};
	\draw (ySIMBA.north) node [above] {SIMBA $\langle y \rangle$};
	\draw (TSIMBA.north) node [above] {SIMBA $\langle T_e \rangle$};
\end{tikzpicture}
\caption{Dependence of spectral distortions on pairs of simulation feedback parameters.
         Each panel shows evaluations of the neural networks trained on the LH set
	 for two astrophysical parameters varied, divided by the simple model
	 in which the dependence factorizes (using the curves from Fig.~\ref{fig:1P}).
	 As can be seen, for IllustrisTNG the factorization approximation is good to
	 within $\sim 5\,\%$, while for SIMBA it is somewhat worse but still reasonable.
	 For readability, the axes ticks have been suppressed. They are exactly identical
	 to the ones in Fig.~\ref{fig:1P}, so that the fiducial model with all feedback
	 parameters equal one is in the center of each panel.}
\label{fig:2P}
\end{figure*}
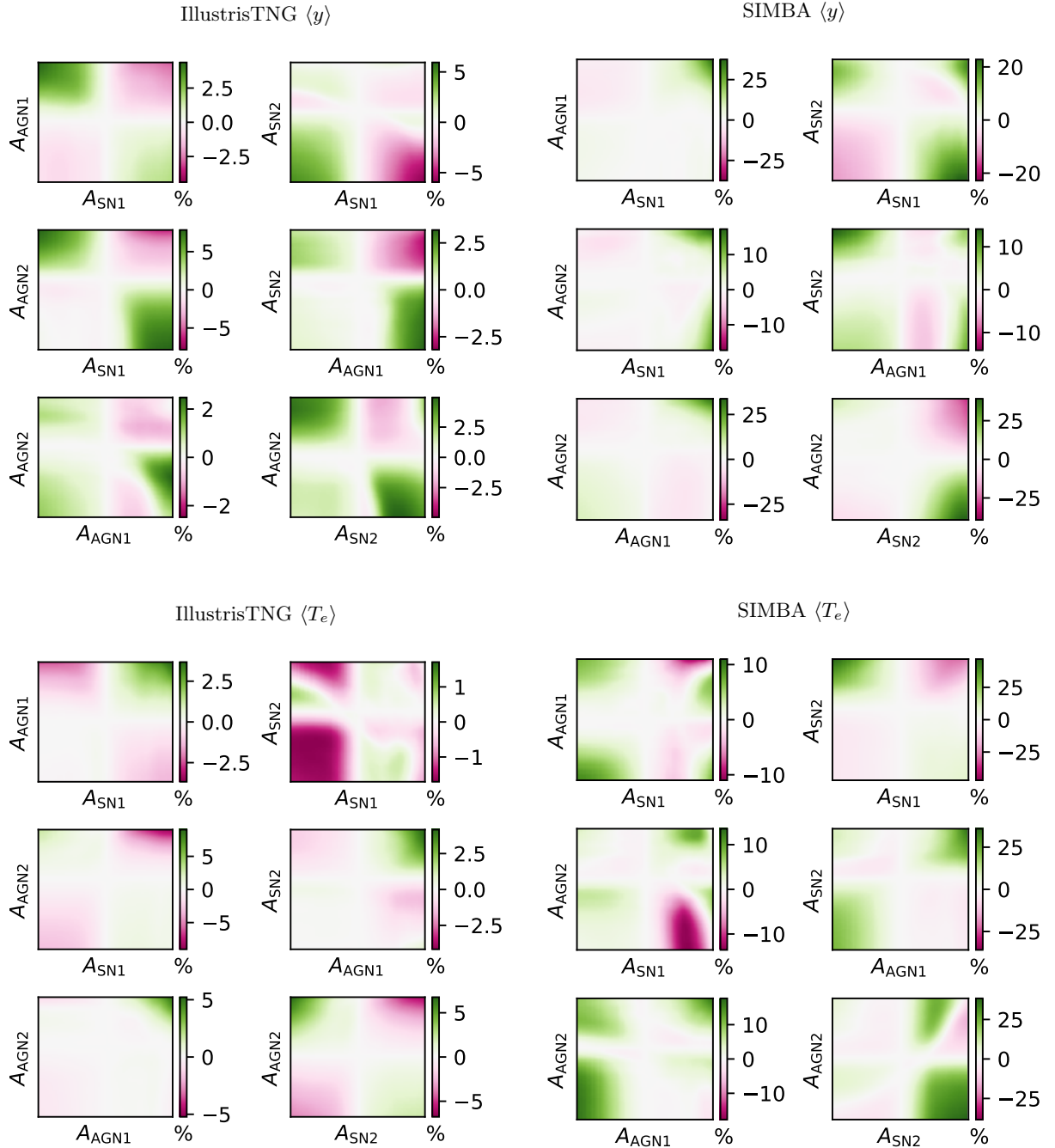

In this appendix, we consider dependence on pairs of parameters.
The primary goal of this exercise is to establish how strongly couplings between feedback parameters
affect the $y$ observables.
In Fig.~\ref{fig:2P}, in each panel we plot the quantity
\begin{equation}
\frac{x_i(A_j, A_k)}{\text{const} \times x_i(A_j) x_i(A_k)} - 1\,,
\end{equation}
where the constant normalizes such that the ratio is one at the fiducial point
and the $x_i$ are evaluations of the neural nets with either one or two parameters
varied from the fiducial point.
Thus, we illustrate deviations from perfect factorization.
We observe that for IllustrisTNG the factorization is a rather good approximation,
with couplings of at most $5\,\%$.
SIMBA exhibits stronger corrections, but the factorization is still relatively
accurate to within $\sim 40\,\%$ and of course the variations are also much larger by about
an order of magnitude.
In units of the overall differences in the $y$ observables the inter-parameters couplings
are quite similar in IllustrisTNG and SIMBA.
It must be noted that the edges of parameter space are likely not quite accurately represented
by the neural nets.
We should emphasize that the neural nets have no structural preference for factorized
representations.
In summary, most of the dependence of $y$ observables on feedback parameters can be readily
read off from Fig.~\ref{fig:1P}.

\section{Symbolic regression}
\label{app:sr}

Symbolic regression identifies equations with parsimonious combinations of input parameters that have the smallest scatter with the given quantity of interest \citep{Wad22,SchLip09, Wadekar2022, WadVil20b, CraSan20,CraXu19, Sha21,UdrTeg20,WuTeg18,KimLu19,LiuTeg11,Wil21}. We employ it to predict $\langle y\rangle$ and $\langle T_e \rangle$ separately as a function of the feedback parameters:
\begin{equation}
\{\langle y\rangle\, ,\, \langle T_e\rangle\} = f(A_\mathrm{SN1},A_\mathrm{AGN1},A_\mathrm{SN2},A_\mathrm{AGN2})\, 
\end{equation}

\begin{figure*}
\includegraphics[width=0.35\textwidth]{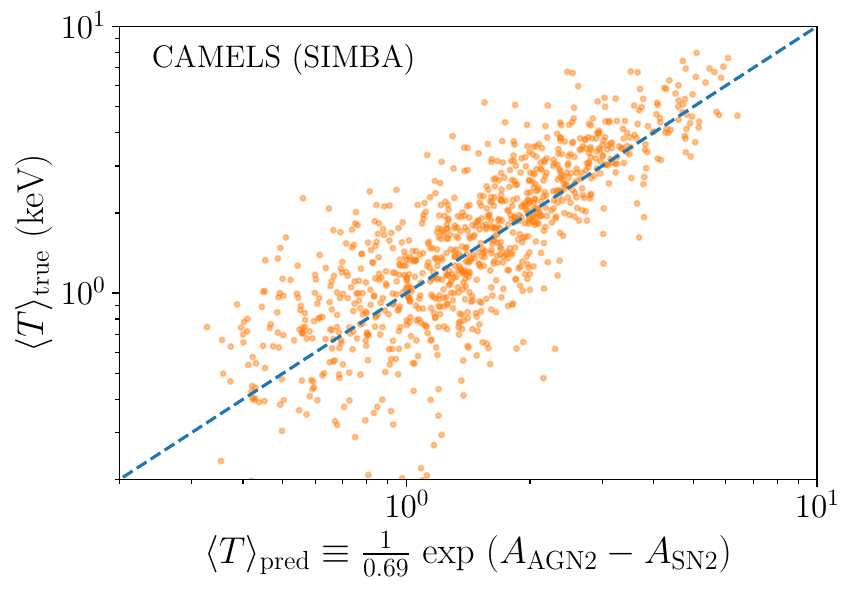}
\includegraphics[width=0.35\textwidth]{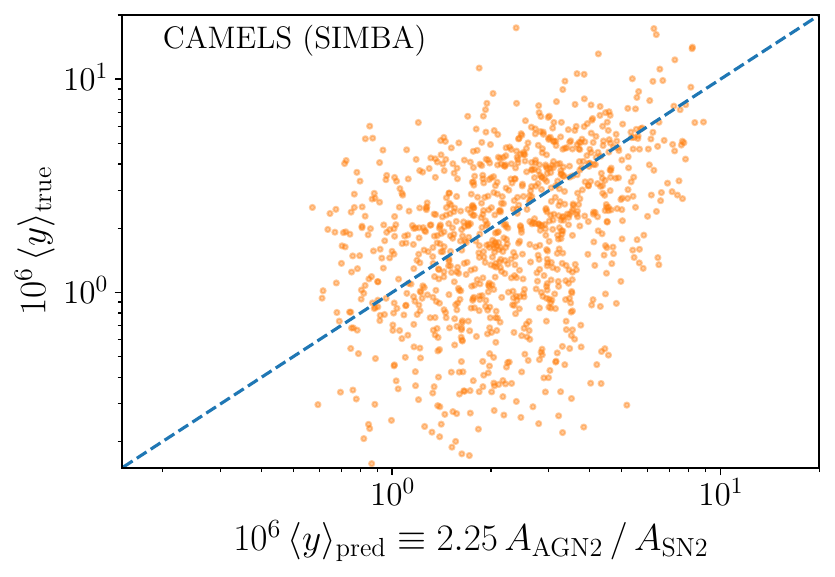}
\includegraphics[width=0.38\textwidth]{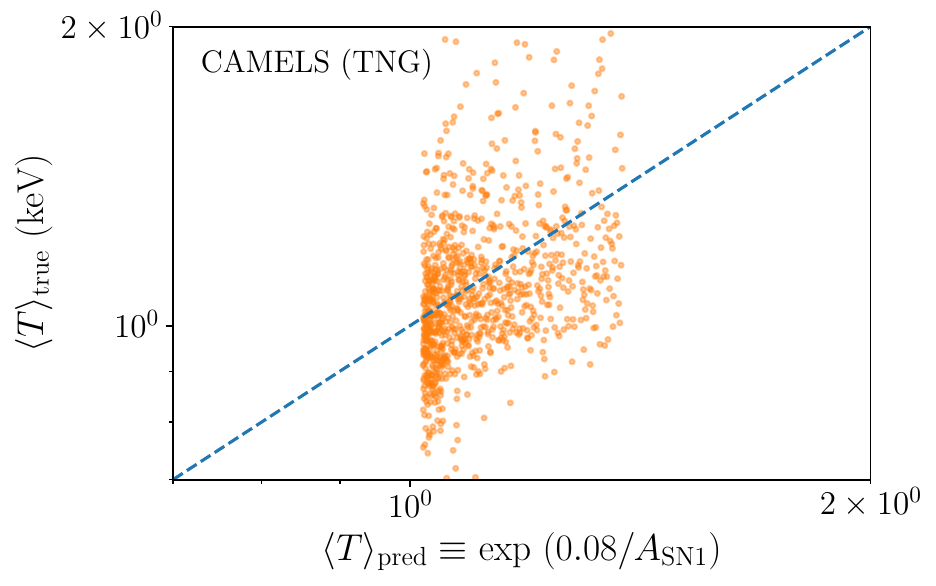}
\includegraphics[width=0.35\textwidth]{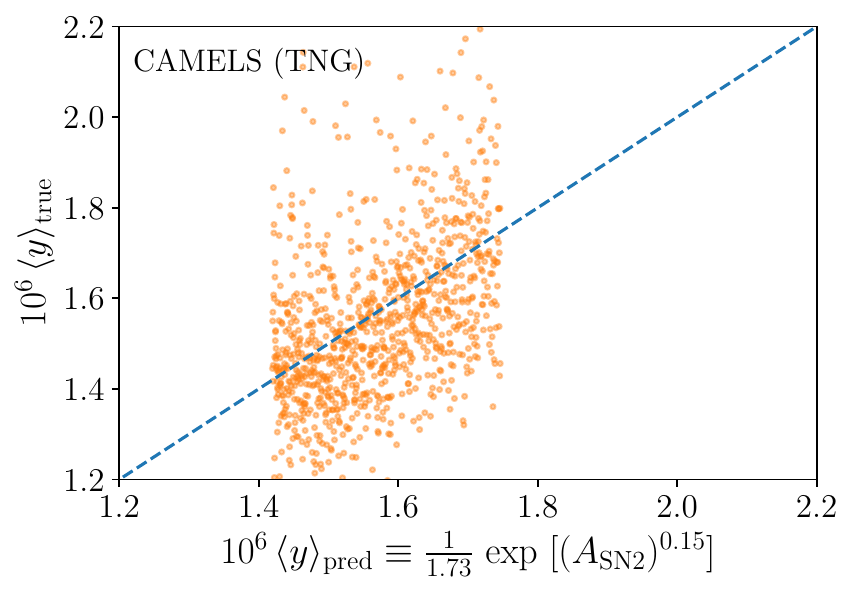}
\caption{As a complement to the results from the neural network, we use a machine learning tool called symbolic regression to predict $\langle y\rangle$ and $\langle T_e \rangle$ separately as an analytic function of the feedback parameters. The blue dashed line represents predicted=true; the closer the points are to the line the more accurate the prediction. The equations perform comparatively better for SIMBA. It is also interesting to see that the effects of AGN and supernovae on both $\langle y\rangle$ and $\langle T_e \rangle$ are opposite to each other for the case of SIMBA.}
\label{fig:SR}
\end{figure*}

We show the equations obtained and their performance in Fig.~\ref{fig:SR}. The data points shown are the measurements from the CAMELS LH set (for the case of TNG, the data has been corrected for sample variance and the cosmology dependence has also been removed, c.f.~Fig.~\ref{fig:hmfcorr}). We also obtained equations more complex than the ones in Fig.~\ref{fig:SR}, however, as the risk of overfitting goes up as the equations get more complex, we show simple ones which have a substantial reduction in the mean squared error. The equations perform much better for the case of SIMBA. This could be either because the dependence of $\langle y\rangle, \langle T_e\rangle$ on the feedback parameters is weaker for TNG, or the dependence for the case of TNG might have a very complicated functional form and it is hard to find a good approximation given the limited size of our data set. Overall, the results in Fig.~\ref{fig:SR} are consistent with the assumption in Eq.~\ref{eq:separable} that the individual parameter feedback dependence can be factorized.

\bibliography{main}{}

\end{document}